\documentclass[a4paper,9pt]{article}

\usepackage{graphicx}
\usepackage{amsmath,amssymb}
\usepackage{setspace}
\usepackage{subcaption}
\usepackage{doi}
\usepackage{float}
\usepackage{multicol}
\usepackage[margin=2.5cm]{geometry}

\title{\bfseries A Comprehensive Diffuse Neutrino Search Using the Full Askaryan Radio Array}

\author{
\textbf{Pawan Giri}\\
for the ARA Collaboration\\[3pt]
University of Nebraska--Lincoln, Lincoln, NE, USA\\
\texttt{pgiri4@huskers.unl.edu}
}

\date{}

\begin{document}

\makeatletter
\renewcommand{\section}{\@startsection{section}{1}{\z@}%
  {0.8ex plus 0.2ex minus 0.1ex}%
  {0.5ex plus 0.1ex minus 0.1ex}%
  {\normalfont\large\bfseries}}
\renewcommand{\subsection}{\@startsection{subsection}{2}{\z@}%
  {0.6ex plus 0.2ex minus 0.1ex}%
  {0.3ex plus 0.1ex minus 0.1ex}%
  {\normalfont\normalsize\bfseries}}
\raggedbottom
\setlength{\parskip}{0pt}
\setlength{\parindent}{1.2em}
\makeatother

\maketitle

\begin{center}
\small
32nd International Symposium on Lepton Photon Interactions at High Energies\\
Madison, Wisconsin, USA, August 25--29, 2025
\end{center}

\vspace{0.6em}

\begin{abstract}
\noindent
The Askaryan Radio Array (ARA) is a neutrino experiment at the South Pole, designed to detect radio-frequency emissions produced by interactions of ultra-high energy (UHE) neutrinos with the Antarctic ice. The array consists of five autonomous stations, each equipped with deep in-ice antennas sensitive to both vertically and horizontally polarized radio signals. With nearly 30 station-years of livetime accumulated, ARA is now conducting its first comprehensive array-wide search for diffuse UHE neutrinos. This analysis is expected to deliver the most stringent constraints from any in-ice radio-based detector up to 1~ZeV and is capable of probing flux levels suggested by KM3NeT around 220~PeV. The results from this analysis marks a critical step toward establishing scalable techniques for next-generation detectors.
\end{abstract}
\clearpage
\vspace{0.8em}

\section{Ultra-high energy neutrinos: origin and motivation}
\begin{figure}[ht]
    \hspace*{-0.8cm}
    \begin{subfigure}[t]{0.50\textwidth}
        \centering
        \includegraphics[height=0.24\textheight]{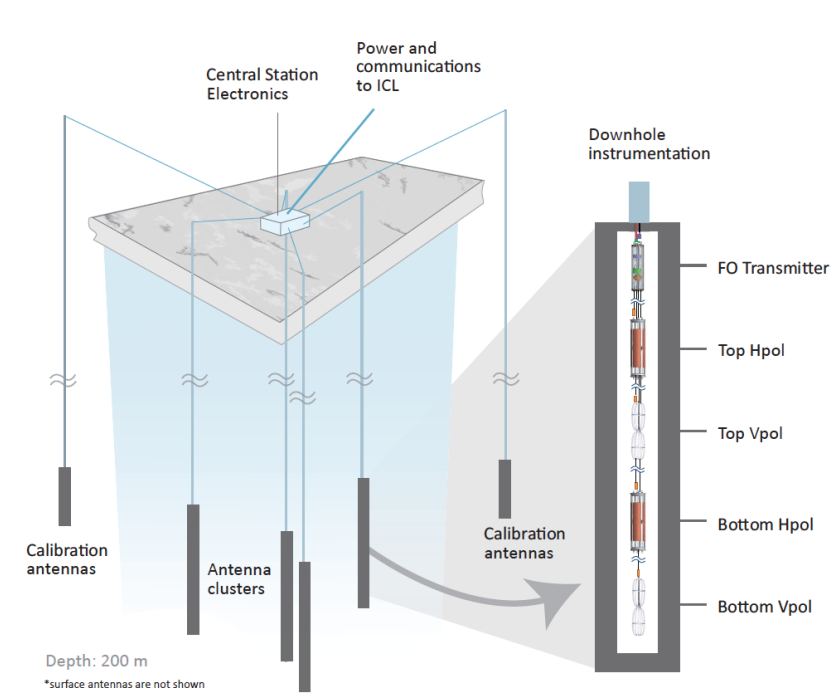}
        \caption{}
    \end{subfigure}
    \hfill
    \begin{subfigure}[t]{0.60\textwidth}
        \centering
        \includegraphics[height=0.22\textheight]{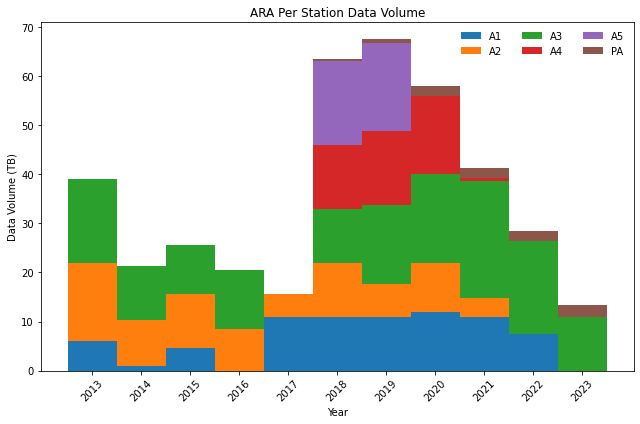}
        \caption{}
    \end{subfigure}
    \caption{(a) Schematic diagram of a traditional ARA station (Stations 1–5), showing antenna geometry and deployment depth. (b) Cumulative data volume collected by all ARA stations during 2013–2023.}
    \label{fig:ara_detector}
\end{figure}

Ultra-high energy (UHE) neutrinos are direct probes of the most energetic particle acceleration processes in the universe. They are produced when UHE cosmic rays interact with ambient photon fields such as the cosmic microwave background and the extragalactic background light. These interactions, primarily through the \(p\gamma\) and \(pp\) channels, generate charged pions that decay into neutrinos, photons, and leptons~\cite{Fiorillo2022}. Because neutrinos interact only weakly and are unaffected by magnetic fields, they can propagate over cosmological distances while preserving both their energy and direction, making them unique messengers of the extreme astrophysical accelerators.

Despite their expected abundance at EeV energies, no diffuse cosmogenic neutrino flux has yet been observed. The IceCube Collaboration currently sets the most stringent upper limits, constraining the proton fraction of ultra-high energy cosmic rays and excluding several optimistic cosmogenic models~\cite{IceCube2025}. These results imply that either the dominant sources evolve weakly with redshift or that heavy nuclei dominate the composition at the highest energies. Recently, the KM3NeT Collaboration reported a 100 PeV-scale neutrino candidate~\cite{KM3NeT2025}. Though below the EeV cosmogenic range, this event provides clear evidence of a neutrino flux at ultrahigh energies.

The extremely low flux and interaction probability of UHE neutrinos require detectors with vast instrumented volumes. Radio detection offers a scalable approach by exploiting coherent Askaryan emission from particle cascades in dense media. The kilometer-scale attenuation length of radio waves in Antarctic ice enables detection over large effective volumes. These advantages motivated the construction of the Askaryan Radio Array (ARA) at the South Pole, extending sensitivity well beyond the PeV regime and enabling exploration of the EeV-scale cosmogenic flux.

\section{Askaryan Radio Array}

ARA is a radio-frequency neutrino detector located near the South Pole, designed to observe nanosecond-scale Askaryan pulses produced by neutrino-induced cascades in ice. Each of the five stations operates independently and contains four vertical strings of broadband antennas deployed to depths of approximately 200~m (Figure~\ref{fig:ara_detector}(a)), covering frequencies between 150 and 850~MHz~\cite{Allison2012}. Station~1 is an exception, with an average antenna depth of about 70~m. Each string hosts both vertically and horizontally polarized antennas, arranged as two pairs located near the bottom of the borehole, with the upper pair positioned about 20~m above the lower pair. This configuration provides polarization sensitivity and the overall station geometry establishes the baselines required for three-dimensional reconstruction of neutrino vertex positions and arrival directions. Station 5 is distinctive in that it includes an additional phased-array string, which gives the station a lower effective trigger threshold than the standard four-string layout.

During operation, each station employs a multi-level trigger and digitization module optimized to capture impulsive signals while suppressing thermal backgrounds. Coincidence logic across multiple antennas further reduces noise, and embedded calibration pulsers continuously monitor timing stability and signal-chain gain.

Analyses of multi-year datasets from the first ARA stations yielded diffuse-flux limits consistent with detailed detector simulations and noise modeling~\cite{Allison2016,Allison2020}. These results demonstrate stable detector performance confirming the feasibility of large-scale radio-based neutrino detection in Antarctic ice. Building on this established performance, the present work extends the search to the full five-station array.


\section{First array-wide neutrino search}

This work presents the ongoing array-wide search for UHE neutrinos using all five ARA stations. The analysis includes data collected between 2013 and 2023 (Figure~\ref{fig:ara_detector}(b)). A unified data analysis framework has been developed to process all stations consistently. This integrated approach improves statistical precision and enables identification of correlated background populations. Applying identical reconstruction and cut definitions across stations minimizes systematic differences and simplifies livetime accounting, producing a coherent dataset suitable for deriving an array-wide flux limit. The analysis is finalizing the optimization of selection cuts using a 10\% blinded subset of data, with full unblinding planned once the final cuts and background estimates are validated.

\subsection{Data cleaning strategy}

For a given station, the events in every run are calibrated and interpolated to uniform time steps. Frequency-dependent phase corrections are applied using the simulated signal-chain response, followed by a 150–850~MHz band-pass filter defining the analysis band. Events affected by readout errors, saturation, or incomplete data are excluded. Runs that coincide with surface activity, calibration pulser operation, or unstable triggers are also removed. A dedicated event-level cut is implemented to remove impulsive background events from cosmic-ray interactions and other anthropogenic sources. Continuous-wave (CW) contamination from anthropogenic and instrumental sources is mitigated using a sine-subtraction algorithm applied per-channel~\cite{CWRemoval}.

\subsection{Detector modeling and noise simulation}

The analysis divides the total dataset from all stations into 39 configurations corresponding to different readout settings and hardware states. For each configuration, noise spectra are obtained from forced triggers and fitted with Rayleigh distributions to reproduce the measured thermal baseline. The fitted spectra are then used to model the frequency-dependent gain of the signal chain and also in the Monte Carlo to generate realistic noise waveforms. Gain and noise responses are derived separately for each antenna, for both vertical and horizontal polarizations. The simulated noise events agree closely with recorded noise data~\cite{5SA}.

\subsection{Neutrino simulation and effective volume estimation}

\begin{figure}[ht]
    \hspace*{-0.8cm}
    \begin{subfigure}[t]{0.47\textwidth}
        \centering
        \includegraphics[height=0.26\textheight]{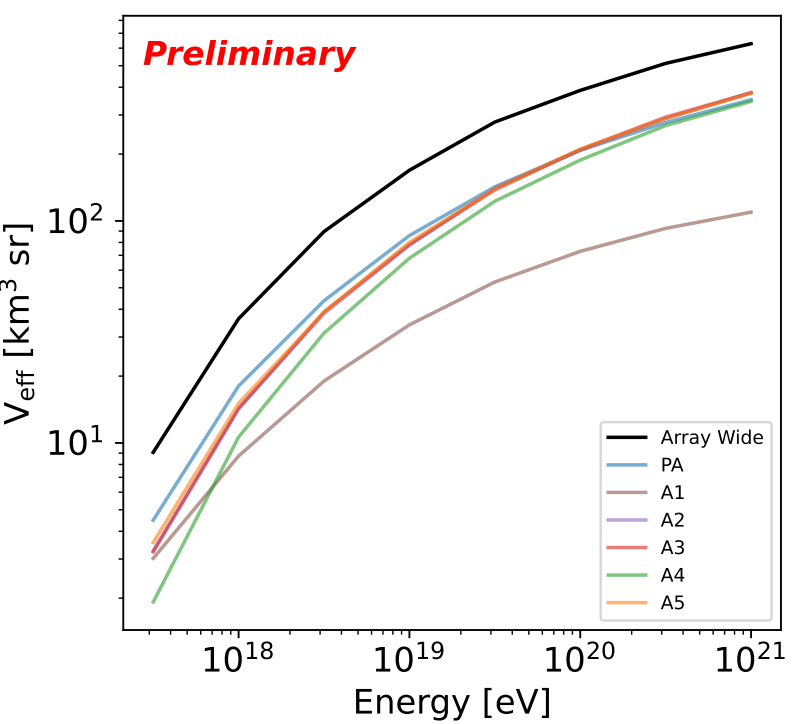}
        \caption{}
    \end{subfigure}
    \hfill
    \begin{subfigure}[t]{0.49\textwidth}
        \centering
        \includegraphics[height=0.27\textheight]{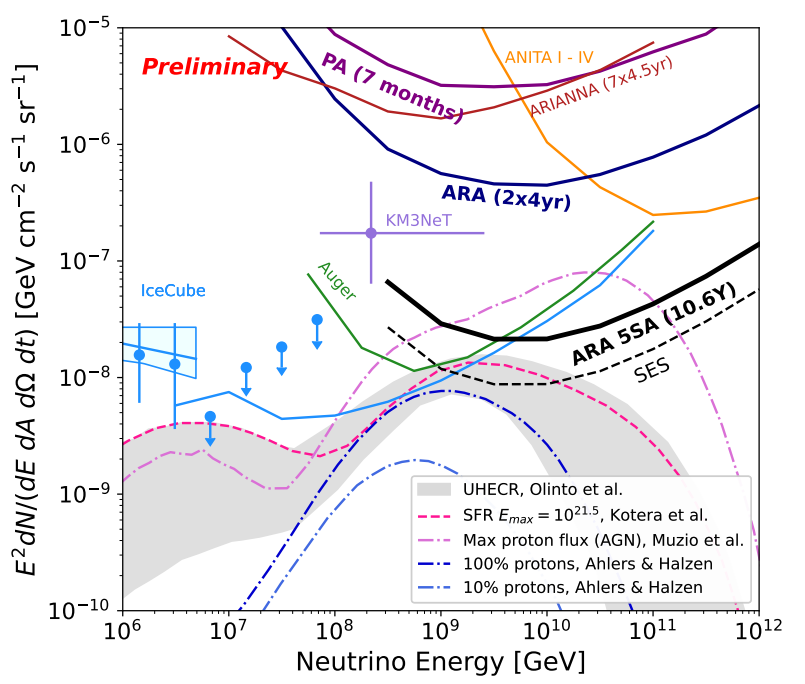}
        \caption{}
    \end{subfigure}
    \caption{(a) Effective volume of the full array compared with individual stations. (b) Projected 90\% C.L. upper limit (solid) and single-event sensitivity (dashed) ~\cite{5SA,Sim}.}
    \label{fig:ara_results}
\end{figure}

Neutrinos of all flavors are generated isotropically within a 15~km radius and 3~km deep cylindrical ice volume centered on the array, including charged and neutral current interactions modeled with standard high-energy cross sections. Askaryan emission from each cascade is computed using semi-empirical parameterizations that include angular dependence, frequency coherence, and shower energy scaling. Signal propagation through the ice follows the measured depth-dependent refractive index profile at the South Pole, and full ray tracing accounts for refraction and attenuation. The simulated electric fields are convolved with the measured antenna and amplifier responses, digitized at 3.2~GSa/s with 12-bit precision, and superimposed with noise sampled from models based on forced trigger data.

The effective volume, $V_{\mathrm{eff}}(E_\nu)$, is obtained from the fraction of simulated neutrinos that trigger the detector and pass all analysis cuts, scaled by the generation volume and solid angle. Livetime weighting for 39 detector configurations yields the combined array exposure for the full dataset. The resulting energy-dependent effective volume for the full array is shown in Fig.~\ref{fig:ara_results}(a), where contributions from individual stations are also indicated. The figure illustrates the gain in acceptance from the array-wide combination compared to single-station analyses. Systematic uncertainties are evaluated by varying the refractive index model, calibration constants, and trigger thresholds within their measured ranges. Secondary cascades from muon and tau tracks, tau decays, and outgoing neutrino interactions, as well as multistation coincidences, are included to increase acceptance at the highest energies. The resulting exposure defines the diffuse flux sensitivity presented in this analysis~\cite{Sim}.


\section{Event selection optimization}

The final stage of this analysis focuses on optimizing the selection of events to maximize sensitivity while constraining backgrounds. The dominant background after earlier cleaning remains thermal noise events, whose rate greatly exceeds any expected neutrino signal. To address this, a linear discriminant (LD) is trained separately for each station and each livetime configuration using cleaned samples of recorded RF noise and simulated neutrino events. The LD assigns every event a score \(t\), and then a cut is applied on \(t\) to remove residual backgrounds while retaining as large a fraction of simulated neutrinos as possible. The optimization is currently ongoing and will be finalized prior to unblinding the full dataset.

The optimization objective is to minimize the expected 90\% confidence-level upper limit on the neutrino flux, \(\phi^{\rm UL}_0\), evaluated for each array-wide configuration \(c\) of stations and years. Formally, one computes  
\[
\phi^{\rm UL}_0 = \frac{\mathrm{FC}\left(\sum_{s} b_{c,s}(t_{c,s})\right)}{4\pi \int \mathrm{d}E\,f(E)\sum_{s}T_c\,A_{\rm eff,c}(E)\,\varepsilon_c(E,t_{c})}\,,
\]
where \(b_{c,s}(t_{c,s})\) is the expected background leakage from station \(s\) in configuration \(c\), \(T_c\) is the livetime, \(A_{\rm eff,c}(E)\) is the effective area for that configuration, \(\varepsilon_c(E,t_{c})\) is the signal efficiency for cut threshold \(t_{c}\), and \(f(E)\) is the assumed neutrino spectrum shape. The FC term refers to the Feldman–Cousins upper limit for the summed background.

By scanning threshold values \(t_{c,s}\) across all configurations spanning the 2013–2023 period, the analysis identifies the combination of cuts that yields the lowest expected flux limit under the chosen spectral assumption. The optimization also accounts for differing station livetimes and effective volumes, ensuring that the final configuration provides an array-wide optimum rather than one biased toward individual stations.

The outcome of the optimization is summarized in Figure ~\ref{fig:ara_results}(b), which presents the projected sensitivity of the full five-station array. The figure shows the expected 90\% confidence-level upper limit on the diffuse neutrino flux and the corresponding single-event sensitivity derived from the optimized selection. The curve reflects the combined livetime and exposure of all configurations and represents the analysis-level performance prior to unblinding. For reference, previous limits from earlier ARA analyses and other experiments are shown together with representative cosmogenic flux predictions. The projection indicates that the full-array search will extend the sensitivity of in-ice radio detectors to energies approaching $1$~ZeV.

The optimized set of LD thresholds, fixed prior to unblinding, maximizes signal efficiency across the array while keeping residual background leakage at a level consistent with a statistically robust flux limit~\cite{5SA}.

\section{Conclusion}

The ARA Collaboration is conducting the first array-wide search for UHE neutrinos using all five traditional stations, encompassing nearly thirty station-years of livetime. The unified calibration, simulation, and analysis framework developed for this work enables consistent treatment across the full array. It also demonstrates the feasibility of performing array-wide searches in radio in-ice detectors. Final analysis of the complete dataset is underway and is expected to yield either candidate UHE neutrino events or the most stringent flux limits from any radio-based experiment to date. This effort establishes the foundation for next-generation detectors such as RNO-G and IceCube-Gen2 Radio.

\begingroup
\scriptsize
\setstretch{0.3}
\clearpage

\begingroup
\scriptsize


\endgroup


\section*{Full Author List: ARA Collaboration (November 07, 2025)}

\noindent
N.~Alden\textsuperscript{1}, 
S.~Ali\textsuperscript{2}, 
P.~Allison\textsuperscript{3}, 
S.~Archambault\textsuperscript{4}, 
J.J.~Beatty\textsuperscript{3}, 
D.Z.~Besson\textsuperscript{2}, 
A.~Bishop\textsuperscript{5}, 
P.~Chen\textsuperscript{6}, 
Y.C.~Chen\textsuperscript{6}, 
Y.-C.~Chen\textsuperscript{6}, 
S.~Chiche\textsuperscript{7}, 
B.A.~Clark\textsuperscript{8}, 
A.~Connolly\textsuperscript{3}, 
K.~Couberly\textsuperscript{2}, 
L.~Cremonesi\textsuperscript{9}, 
A.~Cummings\textsuperscript{10,11,12}, 
P.~Dasgupta\textsuperscript{3}, 
R.~Debolt\textsuperscript{3}, 
S.~de~Kockere\textsuperscript{13}, 
K.D.~de~Vries\textsuperscript{13}, 
C.~Deaconu\textsuperscript{1}, 
M.A.~DuVernois\textsuperscript{5}, 
J.~Flaherty\textsuperscript{3}, 
E.~Friedman\textsuperscript{8}, 
R.~Gaior\textsuperscript{4}, 
P.~Giri\textsuperscript{14$\dagger *$}, 
J.~Hanson\textsuperscript{15}, 
N.~Harty\textsuperscript{16}, 
K.D.~Hoffman\textsuperscript{8}, 
M.-H.~Huang\textsuperscript{6,17}, 
K.~Hughes\textsuperscript{3}, 
A.~Ishihara\textsuperscript{4}, 
A.~Karle\textsuperscript{5}, 
J.L.~Kelley\textsuperscript{5}, 
K.-C.~Kim\textsuperscript{8}, 
M.-C.~Kim\textsuperscript{4}, 
I.~Kravchenko\textsuperscript{14}, 
R.~Krebs\textsuperscript{10,11}, 
C.Y.~Kuo\textsuperscript{6}, 
K.~Kurusu\textsuperscript{4}, 
U.A.~Latif\textsuperscript{13}, 
C.H.~Liu\textsuperscript{14}, 
T.C.~Liu\textsuperscript{6,18}, 
W.~Luszczak\textsuperscript{3}, 
A.~Machtay\textsuperscript{3}, 
K.~Mase\textsuperscript{4}, 
M.S.~Muzio\textsuperscript{5,10,11,12}, 
J.~Nam\textsuperscript{6}, 
R.J.~Nichol\textsuperscript{9}, 
A.~Novikov\textsuperscript{16}, 
A.~Nozdrina\textsuperscript{3}, 
E.~Oberla\textsuperscript{1}, 
C.W.~Pai\textsuperscript{6}, 
Y.~Pan\textsuperscript{16}, 
C.~Pfendner\textsuperscript{19}, 
N.~Punsuebsay\textsuperscript{16}, 
J.~Roth\textsuperscript{16}, 
A.~Salcedo-Gomez\textsuperscript{3}, 
D.~Seckel\textsuperscript{16}, 
M.F.H.~Seikh\textsuperscript{2}, 
Y.-S.~Shiao\textsuperscript{6,20}, 
J.~Stethem\textsuperscript{3}, 
S.C.~Su\textsuperscript{6}, 
S.~Toscano\textsuperscript{7}, 
J.~Torres\textsuperscript{3}, 
J.~Touart\textsuperscript{8}, 
N.~van~Eijndhoven\textsuperscript{13}, 
A.~Vieregg\textsuperscript{1}, 
M.~Vilarino~Fostier\textsuperscript{7}, 
M.-Z.~Wang\textsuperscript{6}, 
S.-H.~Wang\textsuperscript{6}, 
P.~Windischhofer\textsuperscript{1}, 
S.A.~Wissel\textsuperscript{10,11,12}, 
C.~Xie\textsuperscript{9}, 
S.~Yoshida\textsuperscript{4}, 
R.~Young\textsuperscript{2}
\\\\
$^\dagger$Corresponding Authors\\
$^*$Presenter
\\\\
\textsuperscript{1} Dept. of Physics, Dept. of Astronomy and Astrophysics, Enrico Fermi Institute, Kavli Institute for Cosmological Physics, University of Chicago, Chicago, IL 60637\\
\textsuperscript{2} Dept. of Physics and Astronomy, University of Kansas, Lawrence, KS 66045\\
\textsuperscript{3} Dept. of Physics, Center for Cosmology and AstroParticle Physics, The Ohio State University, Columbus, OH 43210\\
\textsuperscript{4} Dept. of Physics, Chiba University, Chiba, Japan\\
\textsuperscript{5} Dept. of Physics, University of Wisconsin-Madison, Madison,  WI 53706\\
\textsuperscript{6} Dept. of Physics, Grad. Inst. of Astrophys., Leung Center for Cosmology and Particle Astrophysics, National Taiwan University, Taipei, Taiwan\\
\textsuperscript{7} Universite Libre de Bruxelles, Science Faculty CP230, B-1050 Brussels, Belgium\\
\textsuperscript{8} Dept. of Physics, University of Maryland, College Park, MD 20742\\
\textsuperscript{9} Dept. of Physics and Astronomy, University College London, London, United Kingdom\\
\textsuperscript{10} Center for Multi-Messenger Astrophysics, Institute for Gravitation and the Cosmos, Pennsylvania State University, University Park, PA 16802\\
\textsuperscript{11} Dept. of Physics, Pennsylvania State University, University Park, PA 16802\\
\textsuperscript{12} Dept. of Astronomy and Astrophysics, Pennsylvania State University, University Park, PA 16802\\
\textsuperscript{13} Vrije Universiteit Brussel, Brussels, Belgium\\
\textsuperscript{14} Dept. of Physics and Astronomy, University of Nebraska, Lincoln, Nebraska 68588\\
\textsuperscript{15} Dept. Physics and Astronomy, Whittier College, Whittier, CA 90602\\
\textsuperscript{16} Dept. of Physics, University of Delaware, Newark, DE 19716\\
\textsuperscript{17} Dept. of Energy Engineering, National United University, Miaoli, Taiwan\\
\textsuperscript{18} Dept. of Applied Physics, National Pingtung University, Pingtung City, Pingtung County 900393, Taiwan\\
\textsuperscript{19} Dept. of Physics and Astronomy, Denison University, Granville, Ohio 43023\\
\textsuperscript{20} National Nano Device Laboratories, Hsinchu 300, Taiwan\\

\vspace{1em}

\section*{Acknowledgements}
\noindent

This analysis was conducted by Abigail Bishop, Brian Clark, Paramita Dasgupta, Pawan Giri, Alan
Salcedo Gomez, Alex Machtay, Marco Muzio, and Mohammad Ful Hossain Seikh. The ARA Collaboration is grateful to support from the National Science Foundation through Award 2013134.
The ARA Collaboration designed, constructed, and now operates the ARA detectors. 
We would like to thank IceCube, and specifically the winterovers for the support in operating the detector. 
Data processing and calibration, Monte Carlo simulations of the detector and of theoretical models and data analyses were performed by a large number
of collaboration members, who also discussed and approved the scientific results presented here. 
We are thankful to Antarctic Support Contractor staff, a Leidos unit for field support and enabling our work on the harshest continent. 
We thank the National Science Foundation (NSF) Office of Polar Programs and Physics Division for funding support. 
We further thank the Taiwan National Science Councils Vanguard Program NSC 92-2628-M-002-09 and the Belgian F.R.S.-FNRS and FWO.
K.~Hughes thanks the NSF for support through the Graduate Research Fellowship Program Award DGE-1746045. 
A.~Connolly thanks the NSF for Award 1806923 and 2209588, and also acknowledges the Ohio Supercomputer Center. 
S.~A.~Wissel thanks the NSF for support through CAREER Award 2033500.
A.~Vieregg, C.~Deaconu, N.~Alden, and P.~Windischhofer thank the NSF for Award 2411662 and the Research Computing Center at the University of Chicago
for computing resources.
R.~Nichol thanks the Leverhulme Trust for their support. 
K.D.~de~Vries is supported by European Research Council under the European Unions Horizon research and innovation program (grant agreement 763 No 805486). 
D.~Besson, I.~Kravchenko, and D.~Seckel thank the NSF for support through the IceCube EPSCoR Initiative (Award ID 2019597). 
M.S.~Muzio thanks the NSF for support through the MPS-Ascend Postdoctoral Fellowship under Award 2138121. 
A.~Bishop thanks the Belgian American Education Foundation for their Graduate Fellowship support.

\end{document}